\documentclass{vgtc}                          % final (conference style)
\ifpdf%                                % if we use pdflatex
  \pdfoutput=1\relax                   % create PDFs from pdfLaTeX
  \usepackage{graphicx}                % allow us to embed graphics files
  \DeclareGraphicsExtensions{.pdf,.png,.jpg,.jpeg} % for pdflatex we expect .pdf, .png, or .jpg files
\else%                                 % else we use pure latex
  \usepackage{graphicx}                % allow us to embed graphics files
  \DeclareGraphicsExtensions{.eps}     % for pure latex we expect eps files
\fi%

\graphicspath{{figures/}{pictures/}{images/}{./}} % where to search for the images

\usepackage{microtype}                 % use micro-typography (slightly more compact, better to read)
\PassOptionsToPackage{warn}{textcomp}  % to address font issues with \textrightarrow
\usepackage{textcomp}                  % use better special symbols
\usepackage{mathptmx}                  % use matching math font
\usepackage{times}                     % we use Times as the main font
\usepackage{cite}                      % needed to automatically sort the references
\usepackage{tabu}                      % only used for the table example
\usepackage{booktabs}                  % only used for the table example
\onlineid{0}

\vgtccategory{Research}

\vgtcinsertpkg

\usepackage{xcolor}

\newenvironment{denseitemize}%
{\begin{itemize}\setlength{\itemsep}{-3pt}\setlength{\parsep}{-3pt}}%
{\end{itemize}}

\title{ASEVis: Visual Exploration of Active System Ensembles\\ to Define Characteristic Measures}
\author{Marina Evers\thanks{e-mail: marina.evers@uni-muenster.de} \\ \scriptsize University of Münster, Germany
\and Raphael Wittkowski\thanks{e-mail: raphael.wittkowski@uni-muenster.de} \\ \scriptsize University of Münster, Germany
\and Lars Linsen\thanks{e-mail: linsen@uni-muenster.de} \\ \scriptsize University of Münster, Germany
}

\teaser{
  \centering
  \includegraphics[width=0.93\linewidth]{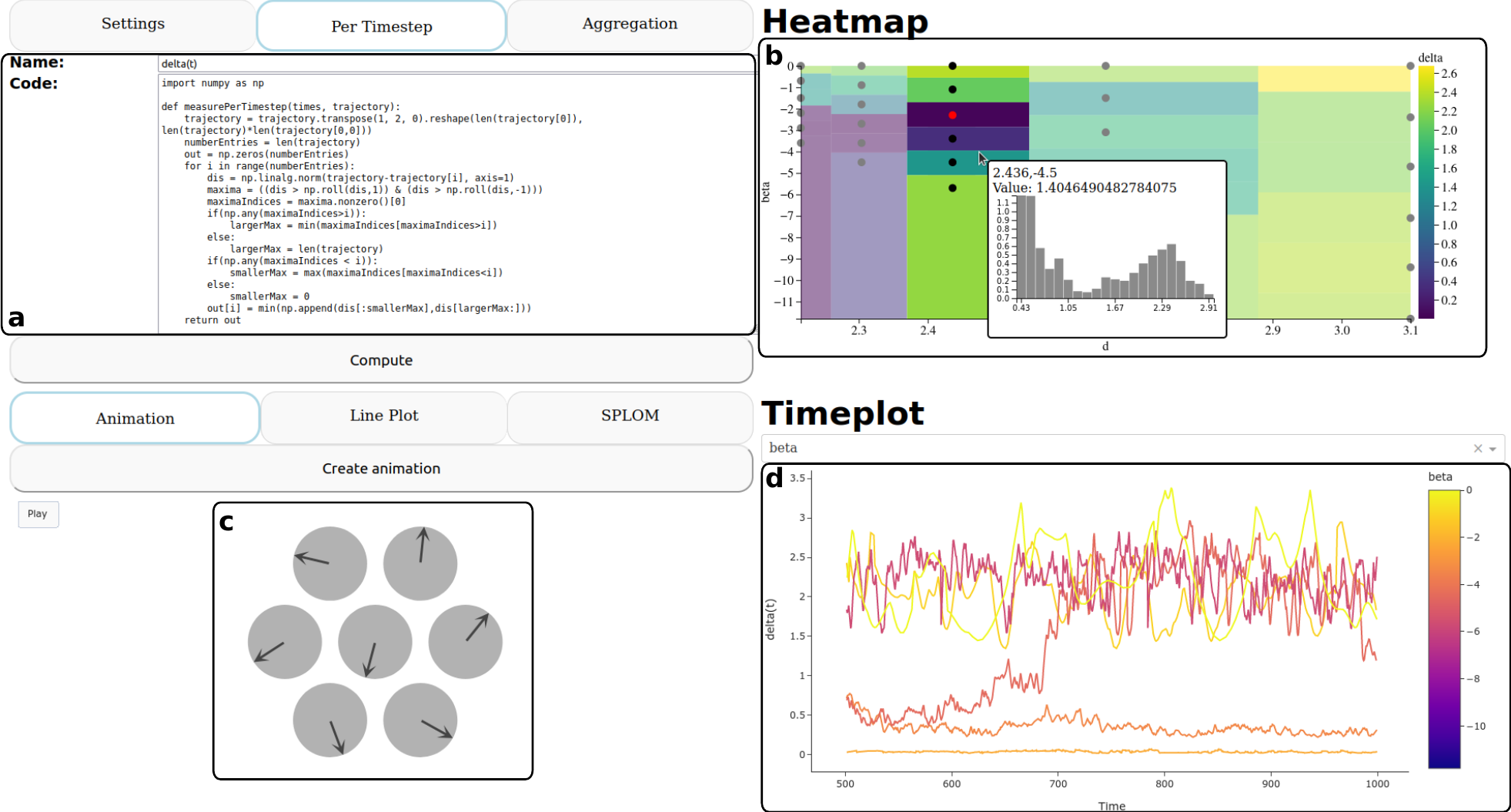}
  \caption{The interactive analysis tool ASEVis: A programming interface (\textbf{a}) allows for the definition of time-dependent measures as well as aggregations. Aggregated measures are shown in a heatmap (\textbf{b}) while the aggregation over time is visualized in the timeplot (\textbf{d}). Detail visualizations for single ensemble members include animations (\textbf{c}), a line plot, and a scatter plot matrix (SPLOM).}
  \label{fig:teaser}
}

\abstract{Simulation ensembles are a common tool in physics for understanding how a model outcome depends on input parameters. We analyze an active particle system, where each particle can use energy from its surroundings to propel itself. A multi-dimensional feature vector containing all particles' motion information can describe the whole system at each time step. The system's behavior strongly depends on input parameters like the propulsion mechanism of the particles. To understand how the time-varying behavior depends on the input parameters, it is necessary to introduce new measures to quantify the difference of the dynamics of the ensemble members. We propose a tool that supports the interactive visual analysis of time-varying feature-vector ensembles. A core component of our tool allows for the interactive definition and refinement of new measures that can then be used to understand the system's behavior and compare the ensemble members. Different visualizations support the user in finding a characteristic measure for the system. By visualizing the user-defined measure, the user can then investigate the parameter dependencies and gain insights into the relationship between input parameters and simulation output.
} % end of abstract

\begin{document}


\firstsection{Introduction}

\maketitle

Numerical simulations are frequently applied in physics because they allow for studying the dependence of a system's behavior on input parameters. One example are \textit{active particles}, which are particles like bacteria or other microorganisms that can propel themselves. Here, simulations are used to identify how the propulsion mechanism of the particles as well as the distances between them influence the system's behavior. The analysis can result in identifying different \textit{states of matter} formed by active particles.

In this article, we want to investigate a so-called active crystal, which is a crystal formed by active particles. Active crystals show interesting properties and might allow for the creation of programmable materials. Each self-propelled particle is fixed in a certain 3D location but can rotate freely in all three dimensions. Considering a small crystal of $k$ particles, where each particle is characterized by a 3-dimensional orientation vector, each state of the system (i.e., each time step of a simulation) can be described by a $3k$-dimensional \textit{feature vector}. Then, the task is to study the evolution of a $3k$-dimensional vector over time and compare it to other ensemble members.
The overall goal is to define a \textit{measure} that describes the evolution of each ensemble member. 
This measure shall characterize the differences in temporal evolutions of the ensemble members.
Many measures exist to describe certain characteristics of 1D time series (such as frequency for a periodic signal), but which measure is most suitable for a given ensemble depends on the dynamics and is a priori unclear. Hence, the temporal evolution of ensemble members needs to be explored in an interactive visual analysis to define (and refine) a suitable measure.
In particular, the measure needs to capture the evolution of all $3k$ dimensions of the feature vector in an aggregated form.
Statistical measures such as mean, median, standard deviation, etc.\ may be applicable to aggregate over dimensions, but again it is unclear a priori which measure is most suitable.
Altogether, the overall measure to be defined shall aggregate over time and dimensions to distinguish the dynamic characteristics of different ensemble members.

We present an interactive visual analysis system to analyze ensembles of multivariate time series, which facilitates the definition and refinement of suitable measures for describing the main dynamics of the system. The visualization of single ensemble members can be used to investigate their dynamical behavior. An interactive programming interface directly embedded into the visual analysis tool then allows for defining measures for capturing the observed dynamical behaviors and reducing the complexity of the data. These measures can be defined for individual time steps or aggregated over time. The definitions can be evaluated interactively, where respective visual representations  allow for comparing different ensemble members and analyzing the dependence of the user-defined measure on the system's input parameters. We evaluate our approach by presenting a case study of how our tool was used to define suitable measures. In fact, by using our tool a new measure was defined, which was used in a recent publication~\cite{evers2021colloidal}. While that article focuses on the results obtained when using the new measure, this paper focuses on the design of the visualization tool and investigates our learnings in the analysis process from a visualization perspective.
%
% Core contribution
Our main contributions can be summarized as follows:
\vspace{-3pt}
\begin{denseitemize}
    \item A requirement analysis and task abstraction for studying complex systems in active particle physics.
    \item A process to interactively define measures that allow for a comparison of ensemble runs, which are described by the evolution of multidimensional feature vectors.
    \item An interactive visual analysis tool that emerges from the requirement analysis and a use case to show how it is used to interactively define a new measure.
\end{denseitemize}

\section{Related Work}
% Ensemble Analysis and parameter space
Recently, a wide range of visualization approaches that focus on different aspects of ensemble data from different domains have been proposed~\cite{Wang2019, Hao2016, potter2009ensemble, sanyal2010noodles}. One key aspect is the analysis of the ensemble's input parameter space~\cite{sedlmair2014visual, bruckner2010result, berger2011parameterspace, evers2022multi} which also motivates our work. For example, Fofonov et al.~\cite{fofonov2018multivisa} proposed a visual analysis approach, where they represent each run as a line and color-code the lines according to the simulations' parameter values. To provide an overview of the parameter space, an important challenge is the definition of derived data~\cite{Landesberger2017, afzal2011Visual, booshehrian2012vismon}. 

% Time series and trajectories
Luboschik et al.~\cite{Luboschik2014, Luboschik2015} focus on the influence of parameters on trajectories. Similar to other works~\cite{nguyen2020visual, obermaier2015visual} they use a set of pre-defined features for the analysis, assuming that the feature of interest is known from the beginning. Zhao et al.~\cite{zhao2011exploratory} propose to create a pipeline that supports creating derived time series interactively, but they do not allow for the definition of aggregations of the data. On the other hand, different approaches tackle the problem of finding features in multidimensional data~\cite{jackle2015temporal, stopar2018streamstory, fujiwara2020visual}, but do not support the derivation of measures used in other visualizations. In general, the majority of work focuses on 2D or 3D trajectories~\cite{he2019variable, lampe2011interactive}. Recently, a method for the generalization to 4D trajectories has been proposed~\cite{amirkhanov2019manylands} as well as dimensionality reductions even to single dimensions~\cite{wulms2021stable}.

% Interactive programming interfaces
Common approaches for the analysis of data with new user-defined measures are provided by interactive notebooks like Jupyter notebook~\cite{jupyter}. It is also possible to use programmable filters in ParaView~\cite{ahrens2005paraview} that allow for deriving data that are then used for visualization. However, interactions in those approaches are limited. Observable notebooks~\cite{observable} support interactions, but being JavaScript-based, they do not allow access to the standard data processing libraries. Furthermore, the notebook-based approaches are very general and do not include ready-to-use visualizations. Cellpackexplorer~\cite{schwarzl2019cellpackexplorer} is a visually aided tool that supports model building but does not target the definition of derived measures. Paraglide~\cite{bergner2013paraglide}, however, proposes an interactive system to analyze the dependence of simulation data on input parameters that can be closely integrated into different programming environments. They identify the construction of derived variables and measures as one requirement for their system, but they do not tackle time-dependent data, which adds an additional layer of complexity. \looseness=-1

\section{Requirement and Task Analysis}
Active matter, which describes systems that consume energy from their surrounding, have raised a lot of interest recently~\cite{BechingerdLLRVV2016}. Nano- or microparticles with this property are called active particles. They use the energy for self-propulsion and are out of equilibrium as all kinds of active matter. This leads to fascinating behavior and exotic properties such as negative viscosities~\cite{saintillan2018rheology}. Active systems have a rich state diagram containing crystalline phases whose properties are not yet fully explored. In this paper, we investigate a small active crystal where the particles are fixed in their positions, as shown in \autoref{fig:teaser}c, but can rotate freely. They interact via hydrodynamic interactions that arise from the effect of their self-propulsion on the surrounding fluid. Thus, the \textit{propulsion mechanism} and the \textit{distance} between the particles determine the interaction and, therefore, the system's characteristic behavior. Numerical simulations are a helpful tool to understand and explore new interesting behavior.

The state of the system at one point in time can be described by a multi-dimensional vector. In our case study, we observe seven particles whose orientation is defined by a 3D vector. Combining these vectors creates a 21-dimensional feature vector containing the complete information about the system's state. The simulations take two input parameters, where one defines the distance between the individual particles (\texttt{d}) and the other defines the propulsion mechanism (\texttt{beta}). Considering the variations over time and the simulation's input parameters, we can characterize the data as an ensemble of multi-dimensional trajectories. A common way in physics to analyze such ensemble data is to use a set of Python scripts or Jupyter notebooks. However, both cases lead to the creation of static plots. Exploring the dataset to find interesting properties and understanding the system's behavior needs many recreations of these graphs, which interrupts the workflow. This is especially true if the analysis includes finding new measures for characterizing and aggregating the data.

Based on our experience within the project described in~\cite{evers2021colloidal}, we defined the following set of requirements for an analysis of the simulation ensemble: \\
\emph{(R1)} To understand the underlying data and how they can be summarized in an expressive way, the \textit{multi-dimensional time series} data for individual simulation runs shall be visually investigated at different levels of detail. \\
\emph{(R2)} Based on the analysis of single ensemble members, it shall be possible to define an \textit{aggregation measure that reduces the dimensionality} of the data. This measure facilitates visualizing the data over time and supports a comparison among different ensemble members. It also allows for determining interesting time intervals for the investigation, e.g., by skipping transition phases. \\
\emph{(R3)} An \textit{aggregation measure over time} shall allow for summarizing the data based on user-defined characteristics. These aggregated values should be compared for the whole ensemble. Additionally, it should be possible to relate the values to the parameter space to see how the user-defined measure varies with changing parameter settings. This approach allows for defining different states of the system, which is a common goal in the analysis of active systems.

From these concrete requirements in the domain of active systems, we can abstract a set of tasks using terms from the field of visualization: Hence, the methods developed can also be applied to other domains that deal with multi-dimensional time-series data: \\
\emph{(T1)} Visualize a single multi-dimensional time series (R1). \\
\emph{(T2)} Interactively define a measure that aggregates the multi-dimensional time series to one scalar value for each time step (R2). \\
\emph{(T3)} Visualize the aggregated multi-dimensional data over time (R2). \\
\emph{(T4)} Interactively define a measure that aggregates a time series to one scalar value (R3). \\
\emph{(T5)} Visualize the aggregated time-series data depending on the parameter values (R3). \\
In the following, we will explain the design choices to adress these tasks (T1-T5) to fulfill the requirements (R1-R3).

\section{Visual Analysis System}

\begin{figure}
\centering
\includegraphics[width=\linewidth]{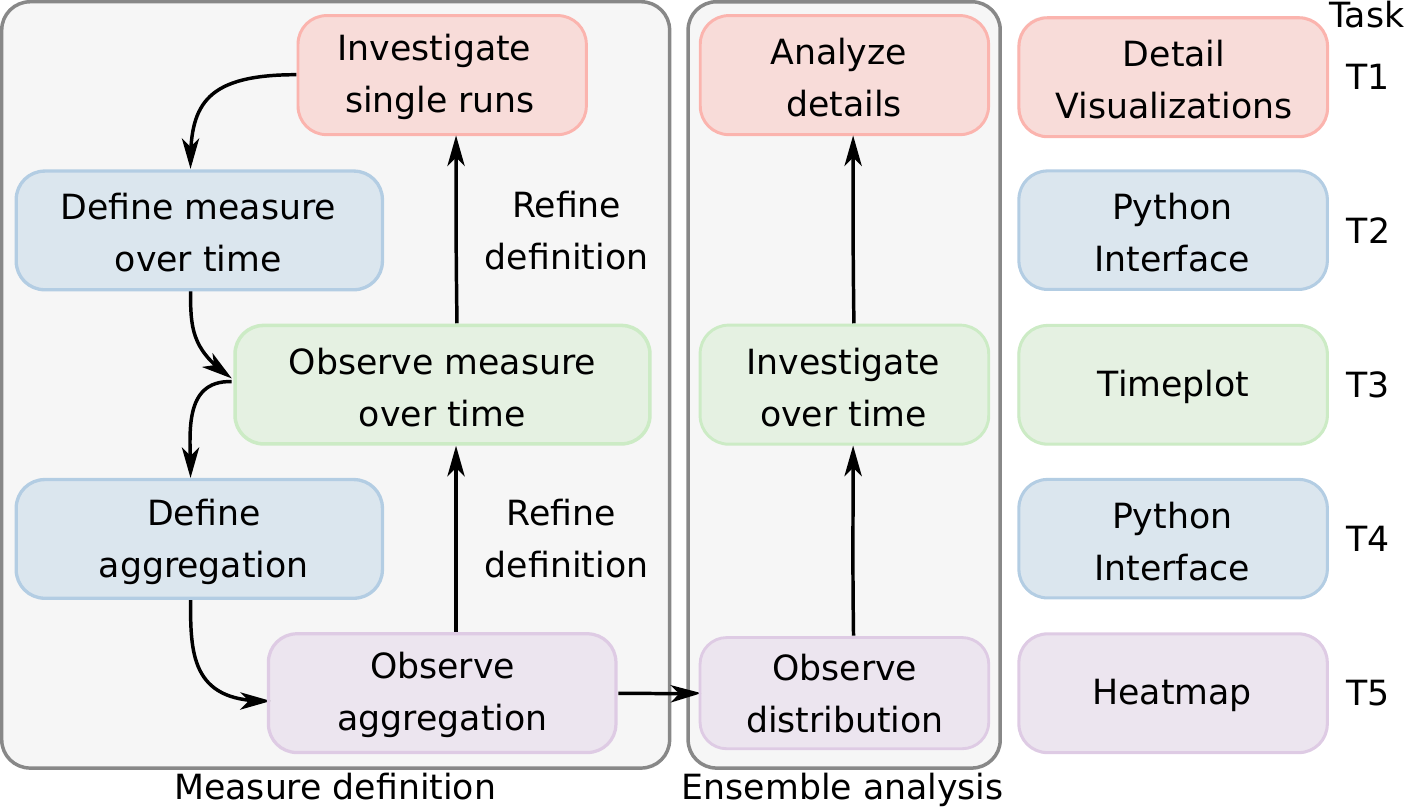}
\caption{Analytical workflow when using ASEVis: Time-dependent and time-independent measures are defined iteratively and then used to gain insights into the data.}
\label{fig:workflow}
\end{figure}

Based on the identified tasks, we designed the interactive analysis system \textit{ASEVis} that combines visualizations at three different levels of detail, see \autoref{fig:teaser}, which we will explain in detail in this section. 
The whole analysis process shown in \autoref{fig:workflow} focuses on an integrated programming interface that allows for the definition of new characteristic measures based on preliminary insights into the data. \looseness-1

A typical analytical workflow starts with a bottom-up analysis: By investigating some simulation runs in detail (T1), the user can identify characteristic behavior to define a measure that aggregates the data for each time step by inserting the definition directly into the interface (T2). This measure is then plotted over time for a selection of ensemble members and shown in a line plot (T3). After some iterative refinement of the time-step measure, if necessary, the user can define a time-series  aggregation measure that results in a single number for each ensemble run (T4). This overall measure is visualized in a heatmap providing an overview of the whole ensemble's dependence on the parameters (T5). 
This bottom-up procedure can be followed by going top-down again.
To start with, the heatmap can interactively be explored by showing the distribution of the values of individual ensemble members that have been used for the aggregation for validation and further insights. 
It is further possible to select a set of runs in the heatmap for which the user-defined measure is shown over time (T3). It is also possible to select a single run that, e.g., shows an interesting behavior based on the user-defined measure. This run can then be investigated in detailed visualizations to understand what causes the extraordinary behavior (T1).

\begin{figure*}[h!]
\centering
\includegraphics[width=1.0\linewidth]{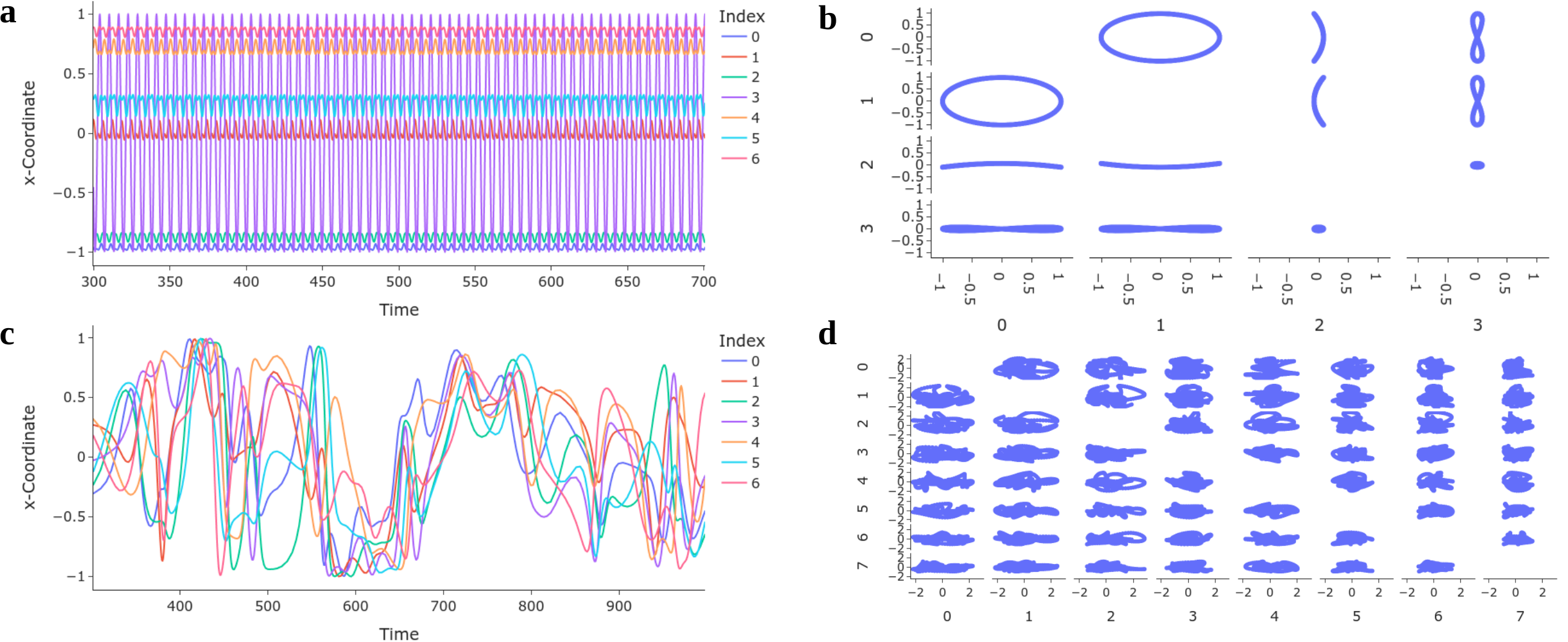}
\caption{Visualizations for the investigation of single runs. \textbf{a} The line plot showing the variation of the x-coordinate for the single particles (\texttt{beta}=$-2.7$, \texttt{d}=$2.3$) indicates a periodic motion. \textbf{b} The SPLOM showing the PCA of the complete data confirms this. \textbf{c} The line plot for the run with \texttt{beta}=$0.0$, \texttt{d}=$2.3$ shows no clear patterns. \textbf{d} The SPLOM also does not reveal a pattern. Here, we need $20$ dimensions to cover $99.9\%$ of the data but show only the first $8$ dimensions.}
\label{fig:details}
\end{figure*}

%\subsection{Detail Visualizations}
\noindent
\textbf{Detail Visualizations} To understand the dynamic behavior of single runs, we combine
three different visualizations. The first one is an animation that directly visualizes the orientation of all particles over time, see \autoref{fig:teaser}c. It provides a descriptive impression of the simulation run that directly shows the system's behavior. However, animations rely on the viewer's memory, and the cognitive demands increase with the number of animations that need to be compared to cover the ensemble structure of the data~\cite{gleicher2011visual}.
Therefore, we also include a visualization of one component (i.e., x-, y-, or z-coordinate) of the individual particle vectors over time. For our system, this results in a line plot with seven lines for the seven particles, as shown in \autoref{fig:details}a and c. Even though this visualization does not capture the complete data but leaves out two components for each particle, it covers the range of features whose identification is more accessible than in an animation.
To cover the whole multi-dimensional trajectory in a static visualization, we include a scatter plot matrix (SPLOM), see \autoref{fig:details}b and d. Instead of directly showing the single coordinates of the multi-dimensional feature, we perform a principal component analysis (PCA)~\cite{jolliffe2016principal} and show the principal components that cover $99.9\%$ of the data (optionally with a maximum number of components), which allows us to identify features in the data by taking the complete information into account. At the same time, the intrinsic dimensionality of the data is identified, which supports finding suitable measures for differentiating regions in parameter space.\looseness-1

\noindent
\textbf{Integration of Measure Definition} For showing aggregated information (T2, T4), we include an interactive programming interface in our analysis tool, see \autoref{fig:teaser}a. It allows users to include the definition of suitable measures directly into the tool and explore the data based on it. We use Python for the programming interface, because it is commonly used in the domain and allows the use of a wide range of libraries like ``numpy'' and ``scipy''. A function template is provided for implementing the measure that can be used as a starting point. The only restriction is given by the function's signature, which needs to be kept. However, due to the clear purpose of the measures, this does not impose restrictions. The interface supports the use of the aggregation measure defined per time step to define the overall aggregation measure. Furthermore, the measures can be assigned a name, which is then used to label the visualizations.

\noindent
\textbf{Comparative Time-dependent Analysis} To allow for a comparative visualization of different ensemble members over time (T3), we include an additional line plot that directly shows the user-defined measure over time, to which we refer in the following as \emph{timeplot}. Here, the measure per time step is evaluated and plotted, see \autoref{fig:teaser}d. To understand how different parameter settings influence the behavior over time, we include the option to color-code the data based on parameter values. The user can interactively switch between color-coding the distance \texttt{d} or the parameter tuning the propulsion mechanism \texttt{beta}.

\noindent
\textbf{Ensemble Overview} An overview of all runs and their dependence on parameters is provided by a heatmap visualization. Here, we chose a heatmap because it resembles a state diagram, which is commonly used in active matter physics. As the parameter space is often sampled irregularly, we fill the gaps by extending the segments around each data point along the axes. However, this encoding makes it hard to identify the exact location of data points. Therefore, we visualize the positions of the samples on top of the heatmap as points. If a run is selected for an analysis in the more detailed views, this point is color coded in red, otherwise in black. These colors do not infer with the perceptually uniform \texttt{viridis} color map, which we chose for encoding the values of the aggregated measure~\cite{mplColormaps}. Selections of regions in the heatmap are shown in full saturation, while the saturation of the unselected cells in the heatmap is decreased to $0.5$. When selecting data points in the heatmap, the temporal evolution of the respective simulation runs is shown in the timeplot. Moreover, when hovering over single cells, a tooltip provides not only the precise numerical values for the measure, but also contains a histogram that shows the distribution of the aggregated data. The histogram allows for a validation of the choice of the aggregation measure as well as additional information about the different runs. \looseness=-1

\section{Use Case}
In the following use case, we showcase how our tool was used to define the order parameter used in the respective study~\cite{evers2021colloidal}. The workflow is also shown in the accompanying video. The dataset contains 30 runs with 465 to 6,664 time steps that cover a time of $1,000R/B_1$ (dimensionless time units). The goal of the simulation ensemble is the study of the influence of the parameters \texttt{d}, which describes the distance between the particles, and \texttt{beta}, which influences their propulsion mechanism. Here, it is of special interest, to see if the systems form different states of matter based on these parameters and where the boundaries of those states of matter are.

% Analyse data with examples of single runs
We start by analyzing individual runs. Here, we can already spot characteristically different behaviors as shown in \autoref{fig:details}. For the run shown in \autoref{fig:details}a, we observe that the systems performs a periodic motion. To confirm if the motion is completely periodic or if this is an artifact by using only part of the data, we visualize the PCA outcome in the SPLOM starting from $300R/B_1$ (see \autoref{fig:details}b). We observe that we only need four principle components to capture $99.9\%$ of the data and that the data indeed exhibit a periodic motion in the 4-dimensional space. However, when investigating a different run, we cannot identify such a periodic motion, neither in the line plot nor in the SPLOM. Thus, the goal is to find a measure that allows us to differentiate these types of motion. This corresponds to a so-called \textit{order parameter} that would also allow to characterize the transition between the states.

In a first step, we tried to use the distance to the first feature vector because this measure would become $0$ every time the system repeats its behavior and thus capture its periodicity. However, by observing this measure for different runs in the timeplot, we found that it does not yield meaningful results for non-periodic cases and strongly depends on the choice of the first time step. Therefore, we adapted our definition and, after some further refinement steps, we came up with a measure that uses  for each timestep the closest Euclidean distance to a position to which the trajectory comes back to. Observing the measure in the timeplot, we found that it allows for a clear separation between different types of runs. Further, it allows us to determine the end of the transition phase to a periodic motion because this measure equals $0$ for periodic cases.

For the aggregation over time, we chose to use the mean over the selected time. The resulting heatmap is shown in \autoref{fig:teaser}b. Here, we can clearly separate the regions with periodic behavior (small values of \texttt{d})
%and around $-3$ for \texttt{beta}
from other regions. We also see a continuous increase in direction of \texttt{d} and discontinuities in direction of \texttt{beta}. Hovering over the heatmap reveals different distributions for the different behaviors. To obtain a better understanding, we can select a set of runs in this direction and investigate it in the timeplot, see \autoref{fig:teaser}d. We also color-code the behavior depending on parameter \texttt{beta} and see that two runs with \texttt{beta}=-2.3 and \texttt{beta}=-3.4 show small values all the time, while those of the other runs exhibit high values (at least partially). The time evolution for \texttt{beta}=-4.5 sticks out, because it shows a jump around $700R/B_1$. To understand this phenomenon, we select this run and analyze it in the animation as well as the other detail visualizations. This reveals that until $700R/B_1$, the motion is close to a periodic pattern, while it seems chaotic for larger times.

\section{Discussion and Conclusion}
We presented an approach for the analysis process in active particle physics. After deriving requirements and abstracting tasks, we proposed a visual system that supports the definition of measures for analyzing complex systems that can be represented by a time-varying multi-dimensional feature vector. Even though our workflow is derived from the domain of active matter physics, it can be generalized to other domains that use ensembles of multi-dimensional trajectories. Only the detail visualizations are  domain-specific. Our approach scales well to larger systems because we use dimensionality reductions by a user-defined measure. However, depending on the number of particles, the detail visualizations should be adapted to avoid overplotting.
Our heatmap visualization is currently limited to analyzing two parameters that suffice for the given application scenario. However, future research directions can target the analysis of higher-dimensional parameter spaces in this context or adapt existing tools~\cite{Luboschik2014}. Additionally, our tool focuses on exploring the data for defining new measures, but is not optimized for creating paper-ready figures targeted at presenting the results, which would improve the practical usefulness of the tool. We provide our source code at \url{https://github.com/marinaevers/asevis}.

\acknowledgments{
This work was funded by the Deutsche Forschungsgemeinschaft (DFG, German Research Foundation) grants 260446826 \mbox{(LI 1530/21-2) and 283183152 (WI 4170/3-2).}}

\bibliographystyle{abbrv-doi}

\bibliography{refs}
\end{document}